\documentclass[aps,preprint,onecolumn,floatfix,nofootinbib,superscriptaddress]{revtex4-1}

\usepackage{amsmath}
\usepackage{amssymb}
\usepackage{amsfonts}
\usepackage{float}
\usepackage{color}
\usepackage{xcolor}
\usepackage{soul}
\usepackage{mathtools}                                  
\usepackage{bbold}                                      
\usepackage{listings}                                   
\usepackage{bm}                                         
\usepackage{subfig}                                     

\newcommand{\rom}[1]{\uppercase\expandafter{\romannumeral #1\relax}}    

\definecolor{ciaccio}{rgb}{2.7,0.3,0.4}
\definecolor{royalblue}{rgb}{0.25, 0.41, 0.88}
\definecolor{pinegreen}{rgb}{0.0, 0.47, 0.44}
\definecolor{alizarin}{rgb}{0.82, 0.1, 0.26}
\definecolor{amarillo}{rgb}{0.95, 0.99, 0.21}
\definecolor{lgray}{gray}{0.70}

\makeatletter
\def\l@subsection#1#2{}
\def\l@subsubsection#1#2{}
\makeatother

\begin{document}

\pagenumbering{arabic}

\title{Explicitly correlated Gaussian functions with shifted-center and projection techniques in pre-Born--Oppenheimer calculations}

\author{Andrea Muolo}
\affiliation{ETH Z\"urich, Laboratory of Physical Chemistry, Vladimir-Prelog-Weg 2, 8093 Z\"urich, Switzerland}

\author{Edit M\'atyus}
\email{correspondence to: matyus@chem.elte.hu}
\affiliation{Institute of Chemistry, E\"otv\"os Lor\'and University, P\'azm\'any P\'eter s\'et\'any 1/A, 1117 Budapest, Hungary}

\author{Markus Reiher}
\email{correspondence to: markus.reiher@phys.chem.ethz.ch}
\affiliation{ETH Z\"urich, Laboratory of Physical Chemistry, Vladimir-Prelog-Weg 2, 8093 Z\"urich, Switzerland}

\vspace*{1.5cm}

\begin{abstract}

Numerical projection methods are elaborated for the calculation of eigenstates of the non-relativistic many-particle Coulomb Hamiltonian 
with selected rotational and parity quantum numbers employing shifted explicitly correlated Gaussian functions, 
which are, in general, not eigenfunctions of the total angular momentum and parity operators. 
The increased computational cost of numerically projecting the basis functions onto the irreducible representations
of the three dimensional rotation-inversion group is the price to pay for the increased flexibility of the basis functions. 
This increased flexibility allowed us to achieve a substantial improvement for the variational upper bound to the 
Pauli-allowed ground-state energy of the H$_3^+=\{$p$^+,$p$^+,$p$^+,$e$^-,$e$^-\}$ molecular ion 
treated as an explicit five-particle system. 
We compare our pre-Born-Oppenheimer result for this molecular ion with rovibrational results
including non-adiabatic corrections.

\end{abstract}

\maketitle

\section{Introduction}\label{SEC:intro}

Energies and wavefunctions of small systems at the low-energy scale can be calculated with very high accuracy. Such calculations
serve as a reference for approximate theories and provide results to compare with high precision experimental measurements.
The continuous advance of experimental techniques as well as theoretical and computational methods enable
scrutinizing expressions for particle interactions and study extensions to the standard model as solutions for puzzling experimental results
of small atoms and molecules, or nuclei and baryons 
\cite{Pachucki2011,Pachucki2013,Pachucki2013_fifth_forces,Korobov2016,Korobov2017,H2+_1,
PhysRevA.86.064502,PhysRevA.79.064502,PhysRevA.77.022509,PhysRevA.87.062506,PhysRevA.97.060501}
Three recent examples are 
(i) relativistic calculations on the H$_2=\{$p$^+,$p$^+,$e$^-,$e$^-\}$ four-particle system \cite{Pachucki2017a}
that demonstrated the effect of the finite size of the proton on the dissociation energy to be $0.000031$ cm$^{-1}$, or $930$ kHz,
(ii) relativistic calculations on the H$_2^+=\{$p$^+,$p$^+,$e$^-\}$ three-particle system \cite{Korobov2017}
that evaluated the energy-level structure with an absolute accuracy of $0.1$ kHz, and
(iii) the transition frequencies of high-$n$ Rydberg states of H$_2$ belonging to a series converging on the ground state
of H$_2^+$ measured with an absolute accuracy of $64$ kHz \cite{Merkt2018}.

Methods for solving quantum-mechanical few-body problems have been employing the family of explicitly correlated Gaussians (ECG) basis functions 
\cite{ECG-history1960,ECG-history1960_1,ECG-history1977,ECG-history1986,ECG-history1993,ECG-history1993,Matyus2011a,Matyus2011b,Adamowicz2013,Adamowicz2013_rev}.
Their application is mostly limited to eigenstates of few-electron atoms and diatomic molecules 
with various total angular momentum values and to natural parity, $p=(-1)^N$. 
A serious difficulty in molecular applications is
obeying the correct rotational symmetry with generally applicable 
$N_p$-particle basis functions. 
For systems for which nonspherical ($N>0$) functions are required, the evaluation of the corresponding matrix elements
becomes increasingly complicated.

The difficulties can be understood by considering the traditional partial-wave construction of the angular part of the basis functions
\begin{equation}
\tilde\theta_{NM_N}(\bm{r}) = 
\left[\left[\left[\mathcal{Y}_{l_1}(\bm{r}_1)\mathcal{Y}_{l_2}(\bm{r}_2)\right]_{N_{12}}\mathcal{Y}_{l_3}(\bm{r}_3)\right]_{N_{123}}\ldots\right]_{N\,M_N}.
\label{eq:veccoup}
\end{equation}
where $l_i$ and $N_i$ are angular momentum quantum numbers and $\mathcal{Y}_{l}(\bm{r})$ are solid spherical harmonics.
The expansion length and, hence, the evaluation time of the matrix elements quickly becomes untractable as the number of particles increases.

Different approaches have been developed in the literature to avoid these difficulties. 
For example, one can restrict the calculation to a special $N$ value and develop the formalism and efficient computer implementations for that case 
\cite{Adamowicz2008_zCECG,Adamowicz2009_N1,Adamowicz2010_N2,Adamowicz2011_N2,Adamowicz2011_N2_Ryd1,Adamowicz2011_N2_Ryd2}. 
Alternatively, the angular motion of the few-body system is described by introducing variationally tunable parameters $u_i$
that depend on the position of the particles and define the so-called ``global vector'' \cite{suzukivarga1998,suzukivarga2008}
\begin{equation}
\label{globalvector}
\bm{v}=\sum_{i=1}^{N_p-1} u_i \bm{r}_i .
\end{equation}

The orbital-rotational motion is then described by the orientation ($\hat{v}=\bm{v}/|\bm{v}|$) of this global vector as follows:
\begin{equation}
\label{GVR}
\theta_{NM_N}\left(\bm{r};\bm{u},K\right)=\left|\bm{v}\right|^{2K+N}Y_{NM_N} (\hat{v}) ~.
\end{equation}

It was shown in Ref.~\cite{suzukivarga1998} that the global vector representation (GVR), Eq.~(\ref{GVR}), 
and the partial wave expansion, Eq.~(\ref{eq:veccoup}), are mathematically equivalent when they are used in a variational procedure 
in which the $u_i$ coefficients are selected based on the energy minimization condition.

In the present work, a direct projection method is developed in which a general (non-symmetric) ECG basis function is projected
onto irreducible representations (irreps) of the three-dimensional rotation-inversion group. 
In the most straightforward application, the basis function parameters are optimized for the non-projected functions. Then, the linear 
variational problem is solved with numerically projected basis functions (without any further non-linear optimization).
Ideally, the parameter optimization would be carried out for the projected functions but this optimal approach is currently limited 
by the computationally expensive task of performing the numerical projection at every iteration of the optimization.
An hybrid approach consisting of iterations on unprojected functions followed by further steps on projected functions 
will be described and employed to obtain substantially improved variational upper bounds for the 
H$_3^+=\{$p$^+,$p$^+,$p$^+,$e$^-,$e$^-\}$ molecular ion as an explicit five-particle system.

\section{Theory}

\section{Translational symmetry}\label{SEC:symmetries}
\begin{figure}[h]
  \centering
  \includegraphics[width=0.30\textwidth]{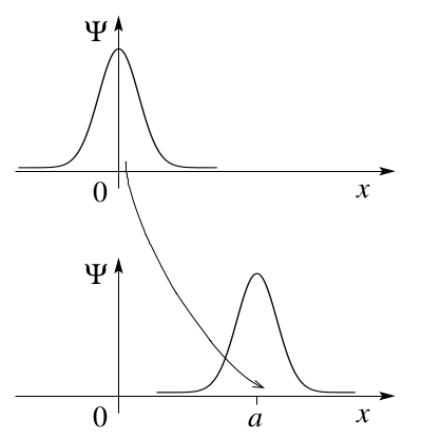}%
  \hfill
  \caption{\label{FIG:symm1}
           Active translation of a wavepacket by $\bm{a}$: $\Psi(x)\rightarrow\Psi(x-a)=\left(U_a\Psi\right)(x)$. }
\end{figure}

Symmetries allow for the classification of solutions and also make their approximation more efficient. 
We briefly introduce basic notation for the following sections on the computational methodology.
The translation of a wavefunction $\Psi\left(\bm{r}\right)$ in the position representation, {\it e.g.} a wave packet at $\bm{r}=0$, 
can be represented by an (active) shift $\bm{a}$ achieved by the operation $\Psi\left(\bm{r}\right)\rightarrow\Psi\left(\bm{r}-\bm{a}\right)$,
see Fig.~(\ref{FIG:symm1}). 

We denote the space translation $\bm{r}\rightarrow\bm{r}+\bm{a}$ by $T_{\bm{a}}\bm{r}=\bm{r}+\bm{a}$ and
the corresponding mapping in the Hilbert space, 
\begin{align}
\left(U_{\bm{a}}\Psi\right)\left(\bm{r}\right) &= \Psi\left(\bm{r}-\bm{a}\right) \nonumber \\
 &{\overset{\bm{a}\,{\text{small}}}{\approx}} \Psi\left(\bm{r}\right) - \bm{a}\cdot\left(\bm{\nabla}\Psi\right)\left(\bm{r}\right) \nonumber \\
 &= \left(\mathbb{1}-{\text{i}}\bm{a}\cdot\bm{p}\right) \Psi\left(\bm{r}\right) \nonumber \\
 &\approx \exp\left(-{\text{i}}\bm{a}\cdot\bm{p}\right) \Psi\left(\bm{r}\right), \label{eq:translations}
\end{align}
produces a representation of the (Abelian) group $G_T=\left\{T_{\bm{a}}|\bm{a}\in\mathbb{R}^3\right\}$ of translation in the
Hilbert space of the wavefunction. We denote this representation by $\mathcal{G}=\left\{U_{\bm{a}}|T_{\bm{a}}\in G_T\right\}$.

The result in Eq.~(\ref{eq:translations}) can be made exact by writing $\Psi\left(\bm{r}-\bm{a}\right)$ 
in a Taylor series or by integrating with an infinitesimal shift $\bm{a}\rightarrow\bm{a}+\delta\bm{a}$, 
$\Psi_{\bm{a}+\delta\bm{a}}=\Psi_{\bm{a}}-\left(i\delta\bm{a}\cdot\bm{p}\right)\Psi_{\bm{a}}$, obtaining the differential
equation $\partial_{\bm{a}}\Psi_{\bm{a}}=-i\left(\bm{p}\right)\Psi_{\bm{a}}$. 
With the initial condition $\Psi_{\bm{0}}=\Psi$ one obtains the solution 
$\Psi_{\bm{a}}=\exp\left(-i\bm{a}\cdot\bm{p}\right) \Psi_{\bm{0}}$.
Hence, the translation operator
\begin{equation}
U_{\bm{a}}=\exp\left(-i\bm{a}\cdot\bm{p}\right)
\end{equation}
describes the active displacement of the wavefunction by $\bm{a}$, where the momentum operator $\bm{p}$ is the
infinitesimal generator of translations. 
With a hermitian $\bm{p}$, $U_{\bm{a}}$ is unitary, $\mathcal{G}$ and $G_T$ are Abelian, continuously connected, isomorphic groups.

\subsection{Spatial rotations and the $SO(3)$ group}

Given an axis of rotation $\bm{\omega}$, and an angle $0\le\omega< 2\pi$, the elements $R_{\bm{\omega}}\in SO(3)$ 
represent 3-dimensional rotations. The corresponding $U(\Omega)$ unitary representation in the Hilbert space of the many particle wavefunctions 
$\Psi\left(\bm{r}\right)$  with $\bm{r}\equiv(\bm{r}_1\ldots\bm{r}_{N_p})^T$ is
\begin{equation}
\left( U_{\bm{\omega}} \Psi \right) \left(\bm{r}\right) = \Psi\left( \left(\mathbb{1}_{N_p}\otimes R_{\bm{\omega}}^{-1}\right) \bm{r}\right).
\end{equation}
 
For small rotations and $N_p=1$ with $R_{\bm{\omega}}^{-1}\bm{r}\sim\bm{r}-\bm{\omega}\wedge\bm{r}$:
\begin{align}
\left( U_{\bm{\omega}} \Psi \right) \left(\bm{r}\right) &\approx \Psi\left(\bm{r}-\bm{\omega}\wedge\bm{r}\right) \nonumber \\
&= \Psi\left(\bm{r}\right) - \epsilon_{ijk} \omega_i r_j \partial_k \Psi\left(\bm{r}\right) + \ldots \nonumber \\
&= \left[\mathbb{1}-{\text{i}}\bm{\omega}\cdot\bm{l}+\ldots\right] \Psi\left(\bm{r}\right) \nonumber \\
&= \exp\left(-{\text{i}}\bm{\omega}\cdot\bm{l}\right) \Psi\left(\bm{r}\right) ,
\end{align}
where the angular momentum operator $\hat{\bm{l}}=-{\text{i}} \bm{r}\wedge\bm{\nabla}$ with
$\hat{\bm{r}}=\left(x,y,z\right)^T$ and $\hat{\bm{\nabla}}=\left(\nabla_x,\nabla_y,\nabla_z\right)^T$.

The representation $U_{\bm{\omega}}=\exp\left(-i\bm{\omega}\cdot\bm{l}\right)$ holds also for large $\bm{\omega}$ rotation angles
and the angular momentum operator is the infinitesimal generator of rotation,
\begin{equation}
U_{\bm{\omega}}=\exp\left(-i\bm{\omega}\cdot\bm{l}\right) \in \mathcal{SO}(3),
\end{equation}
where the group $SO(3)$ and its representation $\mathcal{SO}(3)$ in the Hilbert space $\mathcal{H}$ are isomorphic,
non-Abelian, and continuous.

The orthogonal (or rotation-inversion) group $O(3)$ is the direct product of the special orthogonal group $SO(3)$, 
and the group $C_I=\{\mathcal{E},\mathcal{I}\}$ including the identity and the inversion operator: $O(3)=SO(3)\otimes C_I$.
The irreducible representations of $SO(3)$, labelled with $\mathcal{D}^l$, are $\dim\mathcal{D}^l=2l+1$
dimensional, where $l\in\mathbb{N}_0$.
The irreducible representations of $O(3)$ are $\mathcal{D}^{l\pm}$ with $\dim\mathcal{D}^{l\pm}=2l+1$. 
Spherical harmonics functions, $Y_{lm}$ (of so-called "natural" parity), 
belong to $\mathcal{D}^{l+}$ if $l$ is even and to $\mathcal{D}^{l-}$ if $l$ is odd. 
Functions with unnatural parity can be constructed from combinations of spherical harmonics functions
(see for example \cite{suzukivarga}).

\subsection{Rotations and tensors}

\begin{figure}[h]
  \centering
  \includegraphics[width=0.30\textwidth]{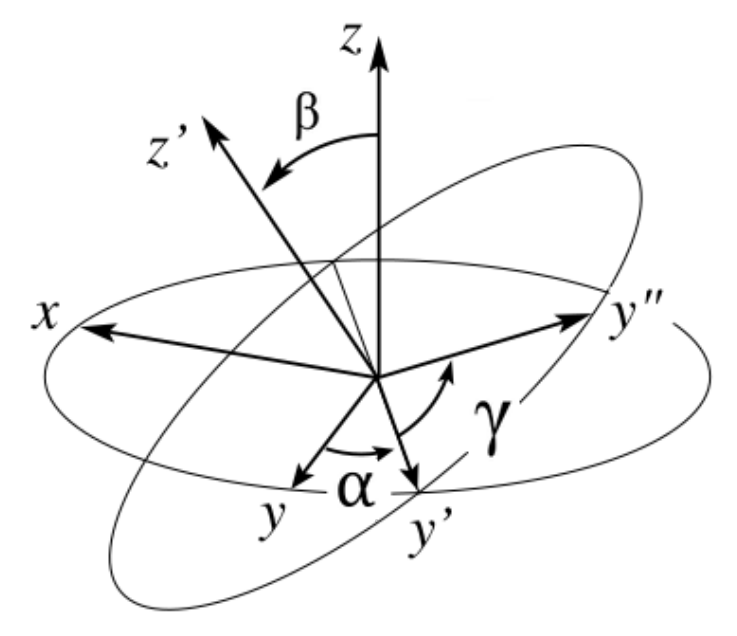}%
  \hfill
  \caption{\label{FIG:rotGroup_3}
           Euler angles: $\alpha$ is a rotation about the $z$ axis and defines $y'$, $\beta$ is a rotation about $y'$,
           and defines $z'$ and $\gamma$ is a rotation about the $z'$. } 
\end{figure}

Let the eigenstate $\left|l,m\right\rangle$ be rotated by $U_{\bm{\omega}}=\exp{\left(-i\bm{\omega}\cdot\bm{l}\right)}$
by an angle $\left|\bm{\omega}\right|$ (positive rotation) about the axis defined by $\hat{\bm{\omega}}$.
An arbitrary rotation operator $\hat{R}$ can then be defined by means of the Euler angles $\alpha,\,\beta,\,\gamma$
(we use the $z-y-z$ convention, see also Fig.~\ref{FIG:rotGroup_3}) as \cite{Rose:AngularMomentum} 
\begin{equation}
\label{RotOperator}
\hat{R}(\Omega) \equiv \hat{R} \left(\alpha,\beta,\gamma\right) = e^{-i\alpha l_z}e^{-i\beta l_y}e^{-i\gamma l_z} ~.
\end{equation}

By definition of an irreducible representation, $\hat{R}(\Omega)$ leaves the irreducible subspace spanned by $\left|l,m'\right\rangle$
with $m'=-l,\ldots,+l$ invariant,
\begin{equation}
\hat{R}(\Omega) \left|l,m\right\rangle = \sum_{m'=-l}^{l} D_{m'm}^{l} ({\Omega}) \left|l,m'\right\rangle,
\end{equation}
where $D_{m'm}^{l}(\Omega)$ is the $(m',m)$-th element of the $l$-th Wigner D-matrix.
Wigner D-matrices define the $(2l+1)$-dimensional irreducible representations of the rotation group, $SO(3)$:
\begin{equation}
\label{WignerDmatrix}
D_{m'm}^{l} ({\Omega}) \equiv \langle l,m' | \hat{R}(\Omega) | l,m \rangle .
\end{equation}

The $D^{l}(\Omega)$ matrices have a number of important properties \cite{Schwabl1}. 
First of all, they are unitary
\begin{equation}
D^{l\dagger}\left({\Omega}\right)D^l\left({\Omega}\right)=D^l\left(-{\Omega}\right)D^l\left({\Omega}\right)=\mathbb{1}
\end{equation}
and also
\begin{equation}
D^{l\dagger}_{m'm}({\Omega}) = D^{l\dagger}_{m\,m'}({\Omega}) = D^l_{m'm}(-{\Omega}).
\end{equation}

Furthermore, the special case  $D_{mm'}^{l} (0,\beta,0) = d_{mm'}^{l}\left(\beta\right)$ is the 'small' d-matrix \cite{Schwabl1}
\begin{align}
d_{mm'}^{l}\left(\beta\right) =& [(j+m')!(j-m')!(j+m)!(j-m)!]^{\frac{1}{2}} \\
&  \times \sum_s \left[\frac{(-1)^s}{(j+m-s)!s!(m'-m+s)!(j-m'-s)!} \right. \\
& \left. \times \left(\cos\frac{\beta}{2}\right)^{2j+m-m'-2s} \left(\sin\frac{\beta}{2}\right)^{m'-m+2s}\right].
\end{align}

Due to their construction, the $D_{m\,m'}^{l}$ elements appear in the rotation expressions of tensor operators. 
An irreducible tensor operator, $T^k$, of order $k$ is transformed by its $2k+1$ components $T_q^k$, 
with $q\in\{-k,-k+1,\ldots,k-1,k\}$ under rotations according to 
\begin{equation}
R(\Omega)\,T_q^k\,R(\Omega)^{-1} = \sum_{q'=-k}^{k} D^k_{q'q}({\Omega}) \, T^k_{q'}.
\end{equation}

\subsection{Hamiltonian and expansion of the wavefunction}\label{SEC:method}

We aim at the variational calculation of the bound states of the many-particle Schr\"odinger Hamiltonian for
$N_p$ particles with Cartesian coordinates $\bm{r}=\left(\bm{r}_1,\ldots,\bm{r}_{N_p}\right)^T$, 
masses $m_i$, and charges $q_i$,
\begin{equation}
\label{nonrel-H}
\hat{H}_{\text{}} = -\bm{\nabla_{r}}^T M \bm{\nabla}_{r} 
+ \sum_{i=1}^{N_p} \sum_{j>i}^{N_p} \frac{q_iq_j}{\left|\bm{r}_i-\bm{r}_j\right|},
\end{equation}
where $\bm{\nabla}_{r}=(\bm{\nabla}_{r_1},\ldots,\bm{\nabla}_{\boldsymbol{r}_{N_p}})^T$ collects the 3-dimensional 
Nabla operators for each particle with 
$\bm{\nabla_{r_1}}=(\frac{\partial}{\partial{r_{1x}}},\frac{\partial}{\partial{r_{1y}}},\frac{\partial}{\partial{r_{1z}}})$.
The entries of the diagonal matrix $M_{ij}=\delta_{ij}\frac{1}{2\,m_i}$ absorb the factor $\frac{1}{2}$.

The many-particle Schr\"odinger Hamiltonian, Eq.~(\ref{nonrel-H}), is invariant to three-dimensional space translation 
and rotation-inversion of the total many-particle system:
\begin{align}
\left[\hat{H}_{\text{}},U_{\bm{a}}\right]&=0 \quad \forall \,\, T_{\bm{a}}\in G_T,\\
\left[\hat{H}_{\text{}},U_{\bm{\omega}}\right]&=0  \quad \forall \,\, R_{\bm{\omega}}\in SO(3),\\
\left[\hat{H}_{\text{}},\hat{I}\right]&=0 \quad \hat{I}\in C_I.
\end{align}
and the operators $\hat{H}_{\text{}}$, $U_{\bm{a}}\in\mathcal{G}_T$, $U_{\bm{\omega}}\in\mathcal{SO}(3)$,
and $\mathcal{I}\in C_I$ have common eigenvectors.
This is an important property, which we would like to build in the basis set in order to design an efficient variational 
procedure for calculating the eigenvalues and eigenfunctions of $\hat{H}_{\text{}}$.

We approximate eigenfunctions of $\hat H_{\text{}}$ in a variational procedure as
\begin{equation}
\label{wavefun}
\Psi(\mathbf{r})=\sum_{I=1}^{N_b}c_{I} \, \bm{\chi}_{I}^{S,M_{S}} \, \hat{Y}\phi_{I}^{{\text{FECG}}}\big(\mathbf{r};\{\omega_I\}\big) ~,
\end{equation}
where the $c_I$ are linear expansion parameters, $\bm{\chi}_{I}^{S,M_{S}}$ the spin functions, 
$\phi_{I}^{\text{{\text{FECG}}}}$ floating explicitly correlated Gaussians (FECGs), and
$\hat{Y}$ is the Young operator projecting onto the appropriate (anti)symmetric subspace.

A basis function $\phi_{I}^{\text{FECG}}$ is defined as 
\begin{align}
\phi_I^{\text{FECG}} \left(\bm{r};A_I^{(r)},\bm{s}_I^{(r)}\right) 
&= \exp\left[-\left(\bm{r}-\bm{s}_I^{(r)}\right)^TA_I^{(r)}\left(\bm{r}-\bm{s}_I^{(r)}\right)\right] 
\label{FECG-fun}
\end{align}
where $A_I^{(r)}=\bar{A}_I^{(r)}\otimes\mathbb{1}_3$ with $\bar{A}_I^{(r)}\in\mathbb{R}^{N_p\times N_p}$ exponents 
and $\bm{s}_I^{(r)}$ positions are optimized variationally.

\subsection{Projection onto O(3) irreducible representations}

A general FECG function, Eq.~(\ref{FECG-fun}), is neither invariant to space inversion nor to space rotation. 
Although this property of the exact eigenfunctions of $\hat H$ is restored in the complete basis set limit, 
this is unfeasible to approach in practice. 
Space translation and the description of the translationally invariant properties have been discussed in detail in our earlier work \cite{Benjamin2013,Muolo2018a},
the results of which are used in the numerical application part of this work.

The broken space-inversion symmetry of an FECG function can however be restored by explicit projection onto the irreps of O(3).

We first consider the $N=0, p=+1$ case (a totally symmetric spherical state), for which the symmetrization
of a general FECG function corresponds to averaging over all possible orientations
\begin{equation}
\label{proj_1}
\phi_I^{{\text{FECG}} \, [N=0]} = \int d\Omega \,\, \hat{R}\left(\Omega\right) \phi_I^{\text{FECG}} \left(\bm{r};A_I^{(r)},\bm{s}_I^{(r)}\right) ~.
\end{equation}
$\hat{R}\left(\Omega\right)$ denotes an active rotation operator and the angular integration is 
\begin{equation}
\int d\Omega \equiv \int_0^{2\pi} d\alpha \int_0^\pi \sin\beta \, d\beta \int_0^{2\pi} d\gamma .
\label{angu_integr}
\end{equation}
By construction $\phi_I^{{\text{FECG}} \, [N=0]}$ is an eigenstate of the square of the total angular momentum 
operator, $\hat{N}^2$ with $N=0$.

How can we now construct functions from FECGs for $N>0$ non-zero angular momentum quantum number?
These functions have a more involved angular node structure. 
In general, the overall rotational symmetry can be recovered by projecting the FECG functions
onto the $N$-th irreducible representation of the rotation group corresponding to the total orbital angular momentum $N$.
We first construct the projection operator, $\hat{P}^{[N]}$, used by Broeckhove and Lathouwers and by several other authors
\cite{numproj_L_theory1,numproj_L_theory,numproj_L_1,numproj_L_2,numproj_L_3},
\begin{equation}
\hat{P}^{[N]} = \sum_{M_N=-N}^{M_N=+N} \hat{P}_{M_NM_N}^{[N]} ,
\end{equation}
with
\begin{equation}
\label{proj_def}
\hat{P}_{M_1M_2}^{[N]} \equiv \int \frac{d\Omega}{4\pi^3} \,\, D^{[N]}_{M_1M_2}\left(\Omega\right)^* \hat{R}\left(\Omega\right) ,
\end{equation}
where $D^{[N]}_{M_1M_2}\left(\Omega\right)$ is the $(M_1 M_2)$-th element of the $N$-th Wigner D-matrix, Eq.~(\ref{WignerDmatrix}) (note the convenient
extension of notation),
and the rotation operators $\hat{R}\left(\Omega\right)$ is expressed in terms of three Euler angles 
$\Omega\equiv\left(\alpha,\beta,\gamma\right)$, Eq.~(\ref{RotOperator}).
Another possible choice for the projection operator is described in Appendix \ref{SEC:alternative_proj_approach}
and follows L\"owdin's idea \cite{newproj_L_0} later reconsidered by Shapiro and Crossley \cite{newproj_L_1,newproj_L_2}

A projected FECG function obtained with the $\hat{P}_{M_1M_2}^{[N]}$ operator, Eq.~(\ref{proj_def}), for $N=0, p=+1$ 
is identical to Eq.~(\ref{proj_1}), which we wrote for simple averaging over all possible orientations
(note that $D^{[0]}_{00}\left(\Omega\right)=1$).

In short, $\hat{P}^{[N]}$ projects any trial function $\varphi(\bm{r})$ onto the eigenspace corresponding to $N$, 
spanned by all eigenfunctions of the $\hat{N}_z$ component of total angular momentum with quantum number
$M_N\in\left[-N,+N\right]$:
\begin{equation}
\varphi^{[N]} \left(\bm{r}\right) = \hat{P}^{[N]} \varphi\left(\bm{r}\right) .
\end{equation}

From the definition in Eq.~(\ref{proj_def}) it follows that $\hat{P}^{[N]}$ is idempotent, Hermitian, and commute 
with $\hat{H}_{\text{}}$ \cite{numproj_L_theory1} due to the rotational invariance of the Hamiltonian:
\begin{align}
P^{[N']}_{M_1M_2}P^{[N]}_{M_3,M_4} =& \delta_{N\,N'}\delta_{M_2M_3}P^{[N]}_{M_1M_4} ,
\label{eq:IdempotencyProjector}
\end{align}
\begin{align}
(\hat{P}^{[N]})^2 = \hat{P}^{[N]}       \hspace{0.3cm},\hspace{0.3cm}
(\hat{P}^{[N]})^\dagger = \hat{P}^{[N]} \hspace{0.3cm},\hspace{0.3cm}
[\hat{H}_{\text{}},\hat{P}^{[N]}] = 0 ~.
\label{eq:PropertiesProjector}
\end{align}

We shall rely on these properties during the calculation of the matrix elements for various quantum mechanical operators.

\subsection{Numerical projection by quadrature}

Given an FECG function and a representation of the $N$-th irrep of the rotation group, we project the FECG function
onto the $M_N$-th subspace by numerically performing the angular integration with Gauss--Legendre quadrature,
\begin{align}
\label{ProjL_numQuad}
\hat{P}_{M_NM_N}^{[N]} \phi_I^{{\text{FECG}}} =& \int \frac{d\Omega}{4\pi^3} 
\, D^{[N]}_{M_NM_N}(\Omega)^* \hat{R}\left(\Omega\right) \phi_I^{{\text{FECG}}} \nonumber \\
\approx& \, \sum_{i_1=1}^n\sum_{i_2=1}^n\sum_{i_3=1}^n 
\omega_{i_1}\omega_{i_2}\omega_{i_3} \, D^{[N]}_{M_NM_N}\left(\Omega_i\right)^* \hat{R}\left(\Omega_i\right) \phi_I^{{\text{FECG}}} ,
\end{align}
with weights 
\begin{equation}
\omega_i = \frac{2\left(1-x_i^2\right)}{\left(n+1\right)^2\left[P_{n+1}\left(x_i\right)\right]^2} ,
\end{equation}
where $x_a\in(-1,+1)$ labels the $a=(1,2,\ldots,n+1)$ roots of the $P_n\left(x\right)$ Legendre polynomials,
and  $\Omega_i=(\alpha_i,\beta_i,\gamma_i)$ are the Euler angles at the quadrature points obtained from scaling the 
$x_a$ points to the appropriate intervals, $\alpha,\gamma\in[0,2\pi)$ and $\beta\in[0,\pi]$.

We rearrange the rotated FECG as
\begin{align}
\hat{R}(\Omega)\phi^{{\text{FECG}}}_I\big(\bm{r};\bar{A}_I^{(r)}\otimes\mathbb{1}_3,\bm{s}_I\big) & = 
\phi^{{\text{FECG}}}_I\big(U(\Omega)^{-1}\,\bm{r};\bar{A}_I^{(r)}\otimes\mathbb{1}_3,\bm{s}_I\big) \nonumber \\
& = \exp\left[-\big(U(\Omega)^{-1}\,\bm{r}-\bm{s}_I\big)^T
\big(\bar{A}_I^{(r)}\otimes\mathbb{1}_3\big)
\big(U(\Omega)^{-1}\,\bm{r}-\bm{s}_I\big)\right] \nonumber \\
& = \exp\left[-\big(\bm{r}-U(\Omega)\bm{s}_I\big)^T
\big(\bar{A}_I^{(r)}\otimes \bar{U}(\Omega)^{-T}\bar{U}(\Omega)^{-1}\big)
\big(\bm{r}-U(\Omega)\bm{s}_I\big)\right] \nonumber \\
& = \phi_I^{{\text{FECG}}}\big(\bm{r};A_I,U(\Omega)\bm{s}_I\big) ~,
\label{eq:FormInvariance}
\end{align}
which means that rotating an FECG in the three-dimensional space is equivalent to a rotation of the shift vector 
defining its center point, $\bm{s}_I\in\mathbb{R}^{3N_p}$. 
It is also important to note that only the parametrization changes ($\bm{s}_I$ is replaced with $U(\Omega)\bm{s}_I$), 
while the mathematical form of the FECG function remains invariant under rotation. 
Employing the ($z-y-z$) convention introduced earlier, the $U(\Omega)$ rotation matrix is obtained from three consecutive in-plane rotations:
\begin{align}
U\left(\Omega\right) &= 
\mathbb{1}_{N_p} \otimes \bar{U}\left(\Omega\right) =
\mathbb{1}_{N_p} \otimes \Big\{U_{\bm{z}}\left(\alpha_i\right)U_{\bm{y}}\left(\beta_j\right)U_{\bm{z}}\left(\gamma_k\right)\Big\} \nonumber \\
&= \mathbb{1}_{N_p} \otimes \left\{
   \left(\begin{array}{ccc} \cos\alpha_i & -\sin\alpha_i & 0 \\ 
                            \sin\alpha_i & \cos\alpha_i & 0 \\ 
                            0 & 0 & 1 \end{array}\right) \cdot 
   \left(\begin{array}{ccc} \cos\beta_j & 0 & -\sin\beta_j \\ 
                                      0 & 1 & 0 \\ 
                            \sin\beta_j & 0 & \cos\beta_j \end{array}\right) \cdot 
   \left(\begin{array}{ccc} \cos\gamma_k & -\sin\gamma_k & 0 \\ 
                            \sin\gamma_k & \cos\gamma_k & 0 \\ 
                            0 & 0 & 1 \end{array}\right) \right\} ~,
\end{align}

where $\mathbb{1}_{Np}$ is indicated in order to emphasize that the entire object is rotated (as a rigid body) 
by $U(\Omega)$ about the origin. 
By exploiting the form invariance of FECGs, Eq.~(\ref{eq:FormInvariance}), 
and the hermiticity and idempotency of $\hat{P}_{M_NM_N}^{[N]}$, Eqs.~(\ref{eq:IdempotencyProjector})--(\ref{eq:PropertiesProjector}),
integrals for a rotationally invariant operator, $\hat{O}$, are evaluated as  
\begin{align}
& \langle\hat{P}_{M_NM_N}^{[N]}\phi_I^{[{\text{FECG}}]}(\bm{r};A_I^{(r)},\bm{s}_I)|\hat{O}|
\hat{P}_{M_NM_N}^{[N]}\phi_J^{[{\text{FECG}}]}(\bm{r};A_J^{(r)},\bm{s}_J)\rangle  \nonumber \\
& \hspace{1.3cm} = \langle\phi_I^{[{\text{FECG}}]}(\bm{r};A_I^{(r)},\bm{s}_I)|\hat{O}|
\hat{P}_{M_NM_N}^{[N]}\phi_J^{[{\text{FECG}}]}(\bm{r};A_J^{(r)},\bm{s}_J)\rangle  \nonumber \\
& \hspace{1.5cm} = \int \frac{d\Omega}{4\pi^3} \, D^{[N]}_{M_NM_N}(\Omega)^* 
\langle\phi_I^{[{\text{FECG}}]}(\bm{r};A_I^{(r)},\bm{s}_I)|\hat{O}|\phi_J^{[{\text{FECG}}]}(\bm{r};A_J^{(r)},U(\Omega)\bm{s}_J)\rangle ~.
\label{eq:ProjMatrixEle} 
\end{align}

\section{Results}

In this section, we explore the projection method for the calculation of rovibrational (rovibronic) states of diatomic systems, 
H$_2^+=\{$p$^+,$p$^+,$e$^-\}$ and H$_2=\{$p$^+,$p$^+,$e$^-,$e$^-\}$, as well as for the triatomic molecular ion H$_3^+=\{$p$^+,$p$^+,$p$^+,$e$^-,$e$^-\}$ 
calculated as three-, four-, and five-particle systems, respectively. 
The Pauli principle is explicitly imposed on both the electrons and the protons in the basis set, Eq.~(\ref{wavefun}). 
We shall label the total proton spin quantum number with $I_p$ and consider singlet (antiparallel, $S_e=0$) electron spin states for H$_2$ and H$_3^+$.

The expectation value of the parity, $\hat{p}$, and the total angular momentum squared operator, 
$\hat{N}^2$, was evaluated to measure the effectiveness of the numerical projection
(the analytical expressions for the $\langle\phi_I^{[{\text{FECG}}]}|\hat{N}^2|\phi_J^{[{\text{FECG}}]}\rangle$ matrix elements
are derived in Appendix \ref{APP:L&LzIntegrals}).
The effect of the overall center-of-mass motion is eliminated during the integral calculations \cite{Muolo2018a} by subtracting
the center-of-mass related terms from the expectation values ({\it{e.g.}} $\langle\hat{H}\rangle$, $\langle\hat{N}^2\rangle$, etc.). 
In Appendix \ref{App:TI_angumom}, we derive the analytical matrix elements for the squared angular momentum operator, the 
projection of the angular momentum onto one axis, and the center of mass elimination expressions.
All $\langle\hat{N}^2\rangle$ and $\langle\hat H\rangle$ in the following tables correspond to translationally invariant expectation values. 

To demonstrate the efficiency of the numerical projection method introduced in this work,
we first build a small basis set composed of only $3$ FECG functions.
This test set is parametrized by converging the first four decimal places of the energy expectation value in a variational procedure.   
Table \ref{TAB:numProj1_H2p_1} collects the results obtained with the projector defined in Eq.~(\ref{proj_def}) 
with $N=0$ and $N=1$ and parity $p=+1$ and $p=-1$, respectively for this small basis set. 
$\langle\hat{N}^2\rangle$ is converged to at least three to four decimal places with 20--50 quadrature 
points for each Euler angle ($\alpha,\beta,\gamma$).

If we start with a set of unprojected functions, 
the plain energy minimization algorithm can build up the ground-state rotational symmetry. 
In other words, the contributions from states of different symmetry is reduced at each iteration (energy-minimization) step.
This observation suggests that the projection after non-linear optimization will perform well for low-$N$ values 
(or high-$N$ states which have a similar internal structure to the low-$N$ states). 

\begin{table*}[h]
\centering
\caption{\label{TAB:numProj1_H2p_1}
         \footnotesize{ Expectation values for $\hat{H}$, $\hat{N}^2$, and $\hat{p}$ in atomic units 
         for a small test set with a fixed number of $N_{\text{b}}=3$ FECG basis functions for the (Pauli-allowed) ground state of 
         H$_2^+=\{$p$^+,$p$^+,$e$^-\}$ with $I_p=0$ and $S_e=\frac{1}{2}$ and 
         H$_3^+=\{$p$^+,$p$^+,$p$^+,$e$^-,$e$^-\}$ with $I_p=\frac{1}{2}$ and $S_e=0$.
         For H$_2^+$ and H$_3^+$ the numerical projection is carried out onto 
         the $(N=0,M_N=0,p=+1)$ and $(N=1,M_N=0,p=-1)$ rotation-inversion functions, respectively. 
         Numerical results are shown for an increasing number of quadrature points $n$ 
         (the same $n$ value is applied for each Euler angle). 
         The $n=0$ case corresponds to unprojected basis functions. }
        }
\footnotesize{
  \begin{tabular} { @{\hspace{1.0mm}} c @{\hspace{3.5mm}} | 
                    @{\hspace{3.5mm}} c @{\hspace{3.5mm}} | 
                    @{\hspace{3.5mm}} c @{\hspace{3.5mm}} c @{\hspace{3.5mm}} c @{\hspace{3.5mm}} | 
                    @{\hspace{3.5mm}} c @{\hspace{3.5mm}} c @{\hspace{3.5mm}} c @{\hspace{1.0mm}} }
  \hline
  \hline
  \multicolumn{5}{c}{\hspace{1.7cm}$N=0$, $M_N=0$, $p=+1$} & \multicolumn{3}{c}{$N=1$, $M_N=0$, $p=-1$} \\
  \hline
  &  $n$   
  &  $\langle\hat{H}\rangle$  &  $\langle\hat{p}\big\rangle$  &  $\langle\hat{N}^2\rangle$ 
  &  $\langle\hat{H}\rangle$  &  $\langle\hat{p}\big\rangle$  &  $\langle\hat{N}^2\rangle$ \\
  \hline
  H$_2^+$ &  $0$   &  $-0.56118$  &  $0.9766$  &  $ +39.018$  &  $-0.55918$  &  $+0.9766$  &  $ +39.018$    \\
          &  $10$  &  $-0.56876$  &  $0.9998$  &  $  -1.114$  &  $-0.56595$  &  $-0.9998$  &  $  +1.074$    \\
          &  $20$  &  $-0.56868$  &  $1.0000$  &  $  -0.002$  &  $-0.56631$  &  $-1.0000$  &  $  +1.997$    \\
          &  $30$  &  $-0.56868$  &  $1.0000$  &  $  +0.000$  &  $-0.56629$  &  $-1.0000$  &  $  +2.000$    \\
          &  $40$  &  $-0.56868$  &  $1.0000$  &  $  +0.000$  &  $-0.56629$  &  $-1.0000$  &  $  +2.000$    \\
  \hline
  H$_3^+$ &  $0$   &  $-1.17164$  &  $-0.0000$  &  $ +86.250$  &  $-1.17164$  &  $-0.0000$  &  $ +86.250$  \\
          &  $10$  &  $-1.14264$  &  $+1.0144$  &  $+124.530$  &  $-1.20678$  &  $-0.9996$  &  $  +5.846$  \\
          &  $20$  &  $-0.96846$  &  $+0.9952$  &  $+104.307$  &  $-1.22468$  &  $-0.9999$  &  $  +3.309$  \\
          &  $30$  &  $-0.84855$  &  $+1.0003$  &  $  +2.317$  &  $-1.22717$  &  $-0.9999$  &  $  +2.063$  \\
          &  $40$  &  $-0.84698$  &  $+1.0000$  &  $  +0.029$  &  $-1.22735$  &  $-1.0000$  &  $  +2.001$  \\
          &  $50$  &  $-0.84697$  &  $+1.0000$  &  $  +0.000$  &  $-1.22735$  &  $-1.0000$  &  $  +2.000$  \\
  \hline
  \hline
  \end{tabular}
}
\end{table*}

Next, we consider a much larger basis set obtained by optimizing 
projected functions directly instead of performing the projection as a separate step. 
In Table \ref{TAB:numProj_H3p_3}, we show results for H$_3^+$ calculated with a moderate basis set size composed of $N_{\text{b}}=120$ FECGs.
The non-linear parameters of the projected basis functions are generated with the competitive selection method 
described by Suzuki and Varga \cite{suzukivarga}, and the selected parameters are refined using Powell's method \cite{Powell04}. 
The projection is carried out onto the $(N=1,M_N=0,p=-1)$ irreducible representation of O(3).
Optimization before projection (Table \ref{TAB:numProj1_H2p_1}) and optimization with projected functions (Table ~\ref{TAB:numProj_H3p_3})
shows that non-linear optimization for projected functions requires
fewer quadrature points to converge the first two to three decimal places of the $\langle\hat{N}^2\rangle$ expectation value. 
Hence, it is computationally less demanding (a smaller number of quadrature points is sufficient) to project a larger,
more tightly pre-optimized basis set to the rotational-inversion irreducible representation of the ground state
($N=0,p=+1$ for H$_2^+$ and $N=1,p=-1$ for H$_3^+$).

\begin{table*}[h]
\centering
\caption{\label{TAB:numProj_H3p_3}
         \footnotesize{ Energy, $\hat{H}$, parity, $\hat{p}$, and squared total angular momentum, $\hat{N}^2$, expectation values 
         in atomic units for H$_3^+=\{$p$^+,$p$^+,$p$^+,$e$^-\,$e$^-\}$ with a total proton spin $I_p=\frac{1}{2}$, 
         and a singlet, $S_e=0$, electronic state obtained with $N_{\text{b}}=120$ optimized FECGs basis functions 
         projected onto ($N=1,M_N=0,p=-1$).
         The expectation values correspond to a growing number of quadrature points, $n$. 
         The $n=0$ row shows results with unprojected basis functions. }
        }
\footnotesize{
  \begin{tabular} { @{\hspace{1.0mm}} c @{\hspace{3.5mm}} | @{\hspace{3.5mm}} c @{\hspace{3.5mm}} c @{\hspace{3.5mm}} c @{\hspace{1.0mm}} }
  \hline
  \hline
          &  $\langle\hat{H}\rangle$  &  $\langle\hat{p}\rangle$  &  $\langle\hat{N}^2\rangle$ \\
  \hline
  $n=0 $  &  $-1.31501$  &  $-0.0000$  &  $+41.233$  \\
  $n=16$  &  $-1.32120$  &  $-1.0001$  &  $ +2.321$  \\
  $n=17$  &  $-1.32123$  &  $-0.9997$  &  $ +2.116$  \\
  $n=18$  &  $-1.32124$  &  $-0.9999$  &  $ +2.091$  \\
  $n=19$  &  $-1.32127$  &  $-0.9999$  &  $ +2.030$  \\
  $n=20$  &  $-1.32115$  &  $-0.9999$  &  $ +2.036$  \\
  $n=21$  &  $-1.32116$  &  $-1.0000$  &  $ +2.014$  \\
  $n=22$  &  $-1.32128$  &  $-1.0000$  &  $ +2.003$  \\
  $n=24$  &  $-1.32128$  &  $-1.0000$  &  $ +2.001$  \\
  \hline
  \hline
  \end{tabular}
}
\end{table*}

Finally, we explore the feasibility of a variational optimization of the numerically projected FECG functions
(np-FECG),
that is optimization after projection.
The non-linear optimization consists of the generation of a good parameter set by competitive selection \cite{suzukivarga}, 
which is refined with Powell's method in repeated cycles.
When performing projection on-the-fly, the variational machinery can generate functions that would require a number of quadrature points 
much higher than $20-25$ (see Table \ref{TAB:numProj1_H2p_1}).
However, since the computational cost of the numerical integration for the Euler angles by quadratures scales exponentially $3^n$,
there are cases for which the quadrature yields matrix elements that are far off the exact value.
We note that we exploit the idempotency of the projector in order to reduce the quadratic scaling with the number of
quadrature points to a linear dependence. However, only the exact projector is strictly idempotent.
If its numerical representation by quadrature was not accurate enough, which also depends on the basis function parametrization,
we had encountered variational collapse and unphysical energies.
We then employed an adaptive quadrature scheme, in which we dynamically adjust the number of points 
and drop trial functions which would require a number of quadrature points above a certain threshold,
to achieve a good compromise between robustness and computational expense.   
This adaptive projection optimization of FECGs remains computationally very demanding and is a practical approach for 
small-sized basis sets (with about $10<N_{\text{b}}<500$). 
Larger basis sets can be handled only if the number of optimization steps per cycle is dramatically reduced.

To calculate tight variational upper bounds which can serve as pre-Born--Oppenheimer (pre-BO) benchmark values for non-adiabatic models,
further improvements in the projection scheme were necessary.
The idea behind Gauss quadrature is to choose $n$ nodes and weights in such a way that polynomials of order $2n+1$
are integrated exactly. The difference between quadratures of order $n$ and $n+1$ can be considered as an error estimate but,
as the zeros of the Legendre polynomials (nodes of the Gaussian quadrature) are never the same for different orders,
$2n+1$ function evaluations must be performed.
As an alternative, we consider the Gauss--Kronrod quadrature \cite{Gauss-Kronrod1},
which is an efficient but nested quadrature scheme.
For the variational optimization we built the basis set with the competitive selection method and
then performed every refining step in the space of projected functions.
Finally the quadrature with respect to the angle $\beta$ was improved by employing Gauss quadrature rules (nodes plus weights)
specifically tailored for the weight function $W(\beta)=\sin\beta$ that is part of the projector operator as shown in Eq.~(\ref{angu_integr}).

\begin{figure}[h]
  \centering
  \includegraphics[width=0.70\textwidth]{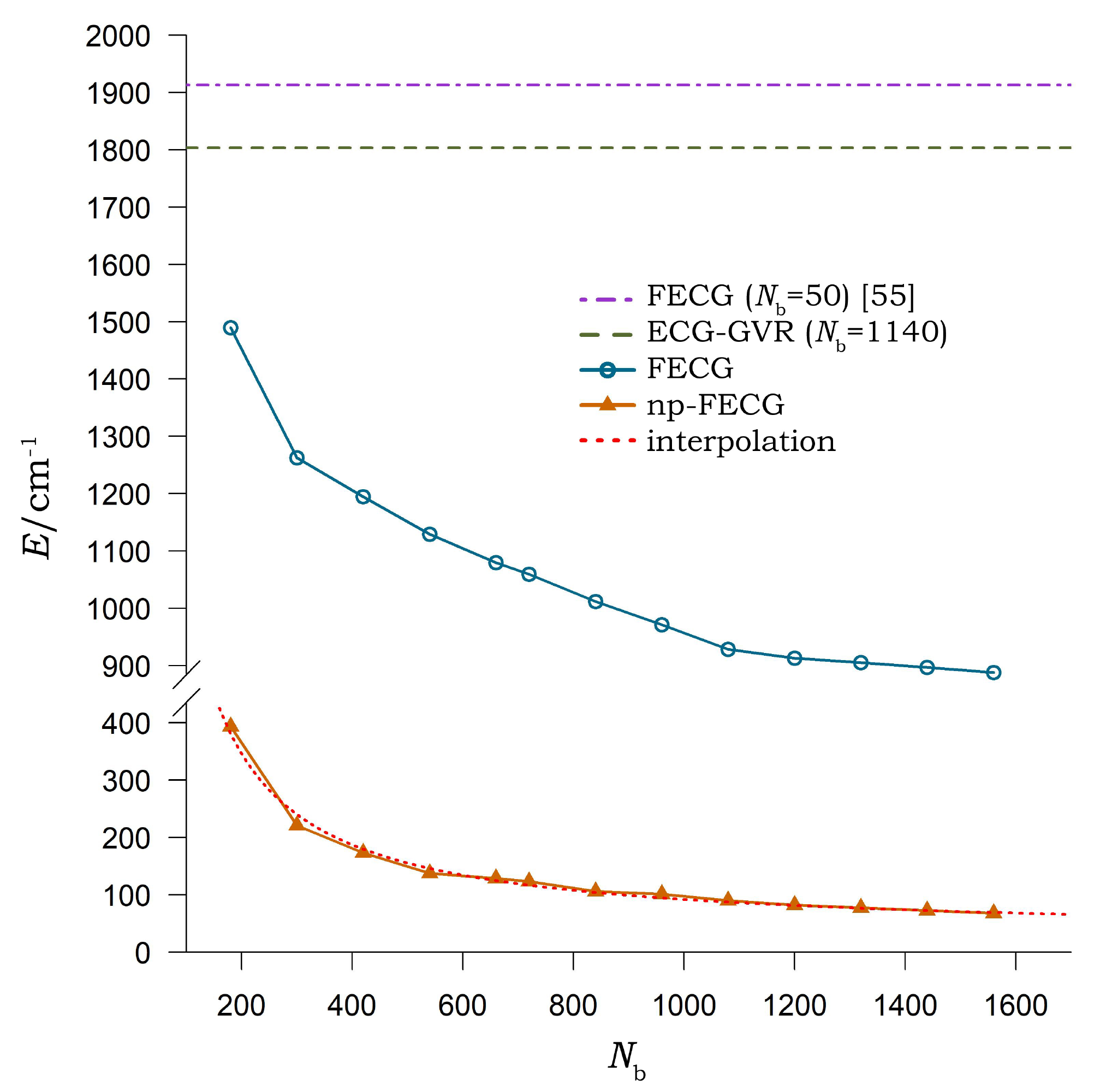}%
  \hfill
  \caption{\label{PLOT:1}
           Energy convergence (in cm$^{-1}$) with respect to the number of basis functions $N_{\text{b}}$, for unprojected and projected basis sets.
           The non-adiabatic estimate was substracted from the energies.
           The dotted line represents the interpolation of the projected FECG energies, while the dashed line shows the best result
           obtained throughout ECG-GVR functions. }
\end{figure}

The energy, parity, and total angular momentum expectation values for both projected and unprojected basis sets 
for optimization after projection are shown in Table \ref{TAB:numProj_H3p_4} and plotted in Fig.~\ref{PLOT:1}.
The lowest-energy pre-BO state with $N=1$, $p=-1$ and $S_p=1/2$ ($S_e=0$) corresponds to the $(J,I_p,p,n)=(1,p,-,1)$ state using 
the notation of Ref.~\cite{LiMc01}.

The data listed in Table \ref{TAB:numProj_H3p_4} allow an extrapolation of the energy to infinite basis-set size.
We considered the inverse power functional form, $E_h=a+b/N_b$, and fitted $a$ and $b$
to the npFECG energies.
The interpolating function is shown in Fig.~\ref{PLOT:1} and the extrapolation to infinite basis-set size 
in Table \ref{TAB:numProj_H3p_4}.

We report also a non-adiabatic estimate (see Table \ref{TAB:numProj_H3p_4}) for the energy of this rotational-vibrational state 
calculated with the GENIUSH program \cite{MaCzCs09,FaMaCs11,MaSzCs14} with the Polyansky--Tennyson model 
(Moss' mass for the vibrations and nuclear mass for rotations) \cite{PoTe99} and the GLH3P potential energy surface \cite{PaAdAl12jcp}.
The GLH3P potential energy surface contains both the diagonal Born--Oppenheimer correction (DBOC) as well as relativistic corrections.
As we consider here the non-relativistic Schr\"odinger Hamiltonian, we removed the relativistic corrections from the potential energy surface
for a proper comparison.
In order to obtain an absolute energy value, we employed the adiabatic electronic energy (BO plus DBOC)
at the equilibrium structure given in Ref.~\cite{CeRyJaKu98}.
This non-adiabatic estimate for the total energy is not variational.
It is based on a perturbative correction to the BO appoximation,
resulting in the diagonal BO correction and non-adiabatic
(mass-correction) effects, which are included here only with a simple
model. Nevertheless, such a set-up is usually considered to be accurate
within about one wavenumber ($<$1 cm-1). A direct comparison of a
variational and (a rigorous) perturbative treatment was recently
presented by Pachucki and Komasa
for rotational states of the four-particle hydrogen molecule \cite{Pachucki2018}. For
the case of H$_3^+$,
the present work represents a significant step toward a variational
validation of effective non-adiabatic models 
for the description of the ground- and, the considerably more
complicated, near-dissociation states; such models may be
developed to compute hundreds and thousands of rovibrational states and
transitions available from experiment.

\begin{table*}[h]
\centering
\caption{\label{TAB:numProj_H3p_4}
         \footnotesize{ Energy, parity, and angular momentum expectation values for the ground states of H$_2=\{$p$^+,$p$^+,$e$^-\,$e$^-\}$ 
         and H$_3^+=\{$p$^+,$p$^+,$p$^+,$e$^-\,$e$^-\}$. $N_{\text{b}}$ is the number of FECGs projected
         onto the $(N=0,M_N=0,p=+1)$ and $(N=1,M_N=0,p=-1)$ angular momentum states.
         The entries in italics represent the best 5-particle variational upper bound for H$_3^+$;
         '$\infty$' denotes the extrapolated result.}
        }
\footnotesize{
  \begin{tabular} { @{\hspace{1.0mm}} c @{\hspace{4.0mm}} | 
                    @{\hspace{4.0mm}} c @{\hspace{4.0mm}} c 
                    @{\hspace{4.0mm}} c @{\hspace{4.0mm}} c 
                    @{\hspace{4.0mm}} c @{\hspace{4.0mm}} c @{\hspace{1.0mm}} }
  \hline 
  \hline
  $N_{\text{b}}$  &  $^a\langle\hat{H}\rangle/{\text E}_{\text h}$  &   
                     $^b\langle\hat{H}\rangle_{\text{proj.}}/{\text E}_{\text h}$    &  
                     $\langle\hat{p}\rangle$  &  $\langle\hat{p}\rangle_{\text{proj.}}$  &
                     $\langle\hat{N}^2\rangle$  &  $\langle\hat{N}^2\rangle_{\text{proj.}}$   \\
  \hline
             &  \multicolumn{6}{c}{\footnotesize H$_2$ ($I_p=0$, $N=0$, $M_N=0$, $p=+1$)} \\
  $440$  &  $-1.162162$  &  $-1.163897$  &  $+0.999955$  &  $+1.000000$  &  $+4.513534$  &  $+0.000054$  \\
  $600$  &  $-1.162358$  &  $-1.163927$  &  $+0.999952$  &  $+1.000000$  &  $+3.976680$  &  $+0.000060$  \\
  $760$  &  $-1.162525$  &  $-1.163945$  &  $+0.999953$  &  $+1.000000$  &  $+3.399123$  &  $+0.000067$  \\
  $920$  &  $-1.162630$  &  $-1.163967$  &  $+0.999939$  &  $+1.000000$  &  $+2.954402$  &  $+0.000050$  \\
  $1080$ &  $-1.162685$  &  $-1.163969$  &  $+0.999945$  &  $+1.000000$  &  $+2.785041$  &  $+0.000024$  \\
  $1240$ &  $-1.162716$  &  $-1.163980$  &  $+0.999952$  &  $+1.000000$  &  $+2.722068$  &  $+0.000052$  \\
  $1400$ &  $-1.162732$  &  $-1.163989$  &  $+0.999944$  &  $+1.000000$  &  $+2.686361$  &  $+0.000052$  \\
  $1560$ &  $-1.162739$  &  $-1.163998$  &  $+0.999946$  &  $+1.000000$  &  $+2.640778$  &  $+0.000053$  \\
  Ref.~\cite{Pachucki2009}  &  $/$  &  $-1.164025$  &  $/$  &  $+1$  &  $/$  &  $0$  \\
  \hline
             &  \multicolumn{6}{c}{\footnotesize H$_3^+$ ($I_p=1/2$, $N=1$, $M_N=0$, $p=-1$)} \\
  $180$    &  $-1.316346$  &  $-1.321340$  &  $-0.000008$  &  $-0.999940$  &  $+36.105414$  &  $+2.001901$  \\
  $420$    &  $-1.317690$  &  $-1.322344$  &  $-0.000174$  &  $-0.999961$  &  $+28.775371$  &  $+2.001383$  \\
  $660$    &  $-1.318216$  &  $-1.322548$  &  $-0.000189$  &  $-0.999980$  &  $+26.623718$  &  $+2.000129$  \\
  $840$    &  $-1.318523$  &  $-1.322652$  &  $-0.000256$  &  $-0.999991$  &  $+24.868238$  &  $+2.000136$  \\
  $1080$   &  $-1.318904$  &  $-1.322726$  &  $-0.000656$  &  $-0.999993$  &  $+22.316282$  &  $+2.000014$  \\
  $1320$   &  $-1.319011$  &  $-1.322782$  &  $-0.001249$  &  $-0.999992$  &  $+21.117237$  &  $+2.000048$  \\
  $1560$   &  $\it -1.319089$  &  $\it -1.322826$  &  $\it -0.001249$  &  $\it -0.999994$  &  $\it +20.843642$  &  $\it +2.000055$  \\
  $\infty$ &  $\it -1.319288$  &  $\it -1.323005$  &  $/$          &  $/$          &  $/$           &  $/$          \\
  $^c$estimate         &  $/$  &  $-1.323146$  &  $/$  &  $(-1)$  &  $/$  &  $(+2)$  \\
  \hline
  \hline
  \end{tabular}
}
\caption*{ \footnotesize{
    $^{a-b}$ $\langle\hat{H}\rangle$ and $\langle\hat{H}\rangle_{\text{proj.}}$ is the lowest eigenvalue of the Hamiltonian 
           obtained with the non-projected and projected FECG basis sets, respectively. 
    $^c$ Non-adiabatic estimate for the energy of this rotational(-vibrational) state. 
         The value of the parity and squared angular momentum for the atomic nuclei are $\langle \hat p_{\text{nuc.}}\rangle=-1$
         and $\langle \hat R^2\rangle=+2$, respectively. These values are indicated in parentheses because they are not strictly 
         the total parity and total angular momentum values of the five-particle system.
} }
\end{table*}

We now compare the results obtained with our numerically projected FECG basis set with results obtained for the ECG-GVR ansatz and from the literature. 
Table \ref{TAB:H3p_ECGGVR} shows the convergence of the (Pauli-allowed) ground-state energy of H$_3^+=\{$p$^+,$p$^+,$p$^+,$e$^-,$e$^-\}$ 
using ECG-GVR basis functions (see Section \ref{SEC:intro}) with $N=1$ and $p=-1$.
Our best result obtained with ECG-GVR is $7.897$ mE$_h$ higher than the best variational upper bound obtained with the numerical projection
method (see Table \ref{TAB:numProj_H3p_4}).
The global vector representation is an excellent alternative of the partial-wave decomposition of the wavefunction.
Its simplicity and generality allowed the calculation of the lowest-energy $N=1$ and $N=2$ states for the $^7$Li atom and
for antiprotonic helium \cite{suzukivarga1998,suzukivarga2008}.
However, the variational reconstruction of the rotational symmetry appears to be cumbersome already for triatomic systems.
The slow convergence of the ground-state energy in Table \ref{TAB:H3p_ECGGVR} shows that the ECG-GVR basis set (with a single global vector)
is impractical for calculating variational upper bounds to the ground-state energy of H$_3^+$.

Earlier results with unprojected FECG functions for the ground-state energy of H$_3^+$
($S_e=0$ and $I_p=\frac{1}{2}$), $E_{p-{\text{H}}_3^+}/{\text E}_{\text h}=-1.314383574$ were reported in Ref.\ \cite{Adamowicz2004},
which is $8.442$ mE$_h$ higher in energy than our best result calculated with numerically projected FECGs (Table \ref{TAB:numProj_H3p_4}).
The numerical projection of FECG functions described in the previous sections
allowed us to substantially improve on the best variational estimates of H$_3^+$.

\begin{table*}[h]
\centering
\caption{\label{TAB:H3p_ECGGVR}
         \footnotesize{ Convergence of the ground state energy of H$_3^+=\{$p$^+,$p$^+,$p$^+,$e$^-\,$e$^-\}$ 
         with $S_e=0$ and $I_p=\frac{1}{2}$ with respect to the number of ECG-GVR functions $N_{\text{b}}$. 
         The general vector representation is employed to describe the natural parity state $(N=1,M_N=0,p=+1)$.
         The $K$ exponent of the polynomial prefactor is randomly selected and optimized from the $\{1,\ldots,20\}$ set. }
        }
\footnotesize{
  \begin{tabular} { @{\hspace{1.0mm}} c @{\hspace{3.5mm}} c @{\hspace{4.5mm}} c | @{\hspace{4.5mm}} c @{\hspace{3.5mm}} c @{\hspace{1.0mm}} }
  \hline
  \hline
  $N_{\text{b}}$   &  $\langle\hat{H}\rangle/{\text E}_{\text h}$  &    &   $N_{\text{b}}$  &  $\langle\hat{H}\rangle/{\text E}_{\text h}$   \\
  \hline
  $ 60$  &  $-1.290269$  &   &  $660$  &  $-1.313773$  \\
  $180$  &  $-1.304466$  &   &  $780$  &  $-1.314465$  \\
  $300$  &  $-1.309775$  &   &  $900$  &  $-1.314716$  \\
  $420$  &  $-1.311816$  &   & $1020$  &  $-1.314850$  \\
  $540$  &  $-1.312467$  &   & $1140$  &  $-1.314929$  \\
  \hline
  \end{tabular}
}
\end{table*}

\section{Conclusions}

The advantage of explicitly correlated Gaussian functions in calculations on 
highly accurate
(non-relativistic) bound states for few-particle systems is due to the analytic and general, $N_{\text{p}}$-particle
integral expressions available for almost all important operators.
Various basis sets with $N\ge0$ total spatial angular momentum quantum numbers (isolated systems) have been proposed in the past.
The traditional partial-wave expansion as well as the more generally applicable (but in variational approaches equivalent)
global vector representations have been used with success. 
Molecular pre-Born--Oppenheimer calculations, especially for systems with more than two heavy nuclei, have turned out to be challenging
because of space rotation-inversion symmetry, correlation effects, and nuclear
motion to be efficiently described simultaneously, including the electronic motion on the same footing.

In this work, we developed numerical projection techniques for the non-symmetric but more flexible basis set of explicitly 
correlated Gaussian functions with shifted centers. The numerical projection ensures the correct spatial rotation-inversion symmetry of the 
variational ansatz, while the shifted ECGs are better suited to describe the (de)localization of the atomic nuclei. 
We presented theoretical as well as technical details for a practical implementation of projected floating ECG functions
in a variational pre-Born--Oppenheimer calculation. 
The first applications of this new numerical approach resulted in an $8.442$ mE$_h$ improvement on the earlier best variational upper bound 
for the (Pauli-allowed) ground-state energy of the H$_3^+=\{$p$^+,$p$^+,$p$^+,$e$^-\,$e$^-\}$ molecular ion
treated as a five-particle system. 
Further possible improvements on the projection approach and the parametrization of the basis set were discussed 
in order to provide five-particle variational benchmark values for selected eigenstates of the H$_3^+$ molecular ion.

\section*{Acknowledgments}

This work has been financially supported by ETH Zurich and the Schweizerischer Nationalfonds (No. SNF\_169120).
EM thanks a PROMYS Grant (No. IZ11Z0\_166525) of the SNSF.

\begin{appendix}

\section{Angular momentum expectation values for FECG functions} \label{APP:L&LzIntegrals}

In this section, we derive matrix elements for FECG functions for the squared total spatial angular momentum operator 
\begin{align}
\hat{{N}^2} = \hat{\bm{N}}_x^2 + \hat{\bm{N}}_y^2 + \hat{\bm{N}}_z^2
\end{align}
with $\hat{\bm{N}}=\sum_{P=1}^{N_p}\hat{\bm{l}}^{(P)}$, which is the sum of angular momentum operators for each particle $P$:
\begin{align}
\hat{\bm{l}}^{(P)} = \hat{\bm{r}}^{(P)} \times \hat{\bm{p}}^{(P)} = -i \left(\hat{\bm{r}}^{(P)}\times\hat{\bm{\nabla}}^{(P)}\right).
\end{align}
With the elementary angular momenta, we re-write the $\hat{\bm{N}^2}$ operator as
\begin{align}
\hat{{N}^2} =& \sum_{P=1}^{N_p} {\left.{\hat{\bm{l}}^{(P)}}\right.}^2 + 2\sum_{P_1<P_2} {{\hat{\bm{l}}}^{(P_1)}}{{\hat{\bm{l}}}^{(P_2)}}  \nonumber \\
=& -\sum_{P=1}^{N_p}\epsilon_{ijk}\epsilon_{ipq} r^{(P)}_{j}\nabla^{(P)}_{k} r^{(P)}_{p}\nabla^{(P)}_{q} -2 \sum_{P_1<P_2} \epsilon_{ijk}\epsilon_{ipq} \,\, r_{j}^{(P_1)}\nabla_k^{(P_1)} r_p^{(P_2)}\nabla_q^{(P_2)}
\end{align}
where the Levi-Civita symbol $\epsilon$ is used together with Einstein summation convention over the $i,j,k,p,q\in\left\{x,y,z\right\}$ indices.
Then, the $i$-th component of $\bm{N}$ is:
\begin{equation}
\hat{N}_i = \sum_{P=1}^{N_p} \hat{l}^{(P)}_i = \frac{1}{i} \epsilon_{ijk} \sum_{P=1}^{N_p} r_j^{(P)}\nabla_k^{(P)}-r_k^{(P)}\nabla_j^{(P)} ~.
\end{equation}
%

\subsection{$\langle\hat{N}_z\rangle$ for FECG functions} \label{APP:LzIntegrals}

The action of $\hat{N}_z$ on FECG functions is given by
\begin{align}
\label{angumomfecg1}
\hat{N}_z\left|\phi\right\rangle =& \frac{1}{i} \sum_{P=1}^{N_p} \left( r_x^{(P)}\nabla_y^{(P)}-r_y^{(P)}\nabla_x^{(P)} \right)
\exp\left[-\bm{s}^TA\bm{s}-\bm{r}A\bm{r}+2\bm{r}^TA\bm{s}\right] \nonumber \\
=& \frac{1}{i} \sum_{P=1}^{N_p} \left( -r_x^{(P)}\bm{A}_{({\text{row}})}^{(P,y)}\bm{r} - \bm{r}^T\bm{A}_{({\text{col}})}^{(P,y)}r_x^{(P)} 
+ 2r_x^{(P)}\bm{A}_{({\text{row}})}^{(P,y)}\bm{s} +r_y^{(P)}\bm{A}^{(P,x)}_{(row)}\bm{r} \right. \nonumber \\
& \left. + \bm{r}\bm{A}_{(\text{col})}^{(P,x)}r_y^{(P)} - 2r_y^{(P)}\bm{A}_{(\text{row})}^{(P,x)}\bm{s} \right) 
\exp\left[-\bm{s}^TA\bm{s}-\bm{r}A\bm{r}+2\bm{r}^TA\bm{s}\right] 
\end{align}
where $\bm{A}_{P,i}^{({\text{row}})}$ ($\bm{A}_{P,i}^{({\text{col}})}$) is the row (column) vector corresponding to the $i$-th 
component of the position vector $\bm{r}_P$.
The pre-exponential terms can be written in a more compact way including the sum over every particle $P$ 
\begin{align}
\label{angumomfecg2}
\sum_{P=1}^{N_p} & \bigg( -r_x^{(P)}\bm{A}_{({\text{row}})}^{(P,y)}\bm{r} - \bm{r}^T\bm{A}_{({\text{col}})}^{(P,y)}r_x^{(P)} + 2r_x^{(P)}\bm{A}_{({\text{row}})}^{(P,y)}\bm{s} +r_y^{(P)}\bm{A}^{(P,x)}_{(row)}\bm{r} + \bm{r}\bm{A}_{(\text{col})}^{(P,x)}r_y^{(P)} - 2r_y^{(P)}\bm{A}_{(\text{row})}^{(P,x)}\bm{s} \bigg) \nonumber \\
=& \bigg[ -\bm{r}^TA^{(y\rightarrow x)}_{({\text{row}})}\bm{r} -\bm{r}^TA^{(y\rightarrow x)}_{({\text{col}})}\bm{r} + 2\bm{r}^TA^{(y\rightarrow x)}_{({\text{row}})}\bm{s} +\bm{r}^TA^{(x\rightarrow y)}_{({\text{row}})}\bm{r} +\bm{r}^TA^{(x\rightarrow y)}_{({\text{col}})}\bm{r} -2\bm{r}^TA^{(x\rightarrow y)}_{({\text{row}})}\bm{s}\bigg] \nonumber \\
=& \bigg[ \bm{r}^T\left( A^{(x\rightarrow y)}_{({\text{row}})} +A^{(x\rightarrow y)}_{({\text{col}})} -A^{(y\rightarrow x)}_{({\text{row}})} -A^{(y\rightarrow x)}_{({\text{col}})} \right)\bm{r} + 2\bm{r}^T\left( A^{(y\rightarrow x)}_{({\text{row}})} -A^{(x\rightarrow y)}_{({\text{row}})} \right) \bm{s} \bigg], \nonumber \\
\end{align}
where $A^{(i\rightarrow j)}_{({\text{row}})}$ $\left(A^{(i\rightarrow j)}_{({\text{col}})}\right)$ is the matrix 
obtained by substituting the row (column) relative to the $i$-th components of every particle position vector 
with the $j$-th component and by setting to zero every other row (column) vector:
\begin{equation}
A^{(y\rightarrow x)}_{({\text{row}})} = 
\left[\mathbb{1}_{N_p\times N_p} \otimes \left(\begin{array}{ccc} 0 & 1 & 0 \\ 0 & 0 & 0 \\ 0 & 0 & 0 \end{array}\right) \right] \cdot A 
\end{equation}
and $A^{(i\rightarrow j)}_{({\text{col}})}=A^{(i\rightarrow j)^T}_{({\text{row}})}$. 

Given the two symmetric matrices $\Omega^{(x,y)}$ and $\omega^{(x,y)}$
\begin{align}
\Omega^{(x,y)} \equiv & A^{(x\rightarrow y)}_{({\text{row}})} +A^{(x\rightarrow y)}_{({\text{col}})} 
-A^{(y\rightarrow x)}_{({\text{row}})} -A^{(y\rightarrow x)}_{({\text{col}})}, \\
\omega^{(x,y)} \equiv & A^{(y\rightarrow x)}_{({\text{row}})} -A^{(x\rightarrow y)}_{({\text{row}})},
\end{align}
it can be seen that, when $A=\bar{A}\otimes\mathbb{1}_3$ is symmetric, $\Omega^{(x,y)}$ is $0$ by construction.
This observation greatly simplifies the calculation of the expectation value calculation $\langle\hat{N}_z\rangle$,
\begin{align}
\label{angumomfecg2.1}
& \left\langle\phi_I\left|\hat{N}_z\right|\phi_J\right\rangle = \frac{2}{i} 
\left\langle\phi_I\left|\bm{r}^T\omega_J^{(x,y)}\bm{s}_J\right|\phi_J\right\rangle ~,
\end{align}
where $A=A_I+A_J$ and the subscript attached to the $\omega$ matrix indicates that the row exchanging 
operation is applied to the $A_J$ correlation matrix belonging to the ket function $\phi_J$.
The integration in Eq.~(\ref{angumomfecg2.1}) is carried out expressing the appropriate 
derivatives of $\phi_J$, which is followed by the evaluation of the standard overlap integral
\begin{align}
&\left\langle\phi_I\left|\hat{N}_z\right|\phi_J\right\rangle = \frac{1}{i} \left(\frac{\partial}{\partial \bm{e}}\omega_J^{(x,y)}\bm{s}_J\right)
\int d\bm{r} \exp\bigg(-\bm{s}^T_IA_I\bm{s}_I-\bm{s}_J^TA_J\bm{s}_J-\bm{r}A\bm{r}+2\bm{r}^T\bm{e}\bigg) \nonumber \\
&= \frac{1}{i} \left(\frac{\partial}{\partial \bm{e}}\omega_J^{(x,y)}\bm{s}_J\right) \left(\frac{\left(2\pi\right)^{3N_p}}{\det\left(A\right)}\right)^{\frac{1}{2}} 
\exp\bigg(-\bm{s}^T_IA_I\bm{s}_I-\bm{s}_J^TA_J\bm{s}_J+\bm{e}^TA^{-1}\bm{e}\bigg) \nonumber \\
&= \frac{2}{i} \left( \bm{e}^TA^{-1}\omega_J^{(x,y)}\bm{s}_J \right) \left\langle\phi_I|\phi_j\right\rangle ~,
\label{eq:APP:LzIntegral:FECG-overlap-integral}
\end{align}
with $A=A_I+A_J$ and $\bm{e}=A_I\bm{s}_I+A_J\bm{s}_J$. For the diagonal matrix elements, one finds
\begin{align}
\left\langle\phi_I\left|\hat{N}_z\right|\phi_I\right\rangle = \frac{2}{i} \left( \bm{s}^T\left[
\mathcal{A}\otimes \left(\begin{array}{ccc} 0 & 1 & 0 \\ 0 & 0 & 0 \\ 0 & 0 & 0 \end{array}\right) -
\mathcal{A}\otimes \left(\begin{array}{ccc} 0 & 0 & 0 \\ 1 & 0 & 0 \\ 0 & 0 & 0 \end{array}\right) \right] \bm{s} \right) 
\left\langle\phi_I|\phi_j\right\rangle = 0.
\end{align}

\subsection{$\langle\hat{{N}}^2\rangle$ for FECG} \label{APP:L2-FECG}

First, we re-write the integrals by exploiting the Hermiticity of $\hat{N}$ as 
\begin{equation}
\label{angumomfecg4}
\langle\phi_I|\hat{N}^2|\phi_J\rangle = \langle\hat{\bm{N}}\phi_I|\hat{\bm{N}}\phi_J\rangle.
\end{equation}
The terms originating from the action of the $\hat{N}_i=\sum_{P=1}^{N_p}{l}^{(P)}_i$ on the bra and the ket 
functions have already been derived in Section \ref{APP:LzIntegrals},
\begin{align}
\langle\hat{N}_i\phi_I| = - \frac{2}{i}\epsilon_{ipq}' & \langle\phi_I| \big[ \bm{r}^T\omega_{I}^{(p,q)}\bm{s}_I \big] \label{angumomfecg4.1}
\end{align}
and
\begin{align}
|\hat{N}_i\phi_J\rangle = + \frac{2}{i}\epsilon_{ijk}' & \big[ \bm{r}^T\omega_{J}^{(j,k)}\bm{s}_J \big] |\phi_J\rangle, \label{angumomfecg4.2}
\end{align}
where $\epsilon_{abc}'$ is non-zero only when the corresponding Levi-Civita symbol is equal to $+1$ and
$\bm{A}_{I,P_1,k}^{\text{(row)}}$ is a row vector of the $A_I$ matrix: specifially, the row associated with the 
$k$-th component of the $P_1$ particle position vector.
Then, the integral in Eq.~(\ref{angumomfecg4}) becomes 
\begin{align}
\langle\hat{\bm{N}}\phi_I|\hat{\bm{N}}\phi_J\rangle =& \int \epsilon_{ijk}' \left[ 4\left(\bm{r}^T\omega_{I}^{(j,k)}\bm{s}_I\right) \left(\bm{r}^T\omega_{J}^{(j,k)}\bm{s}_J\right) \right] \nonumber \\
& \times \exp\left( \bm{s}^T_IA_I\bm{s}_I - \bm{s}_J^TA_J\bm{s}_J - \bm{r}^T\underset{\equiv A}{\underbrace{\left(A_I+A_J\right)}}\bm{r} + 2\bm{r}^T\underset{\equiv\bm{e}}{\underbrace{\left(A_I\bm{s}_I+A_J\bm{s}_J\right)}} \right) d\bm{r}.
\end{align}
We collect the pre-exponential terms from the integration by writing them in terms of derivatives of $A$ and $\bm{e}$.
The remaining integrand function is that of the simple $\langle\phi_I|\phi_J\rangle$ overlap integral given 
in Eq.~(\ref{eq:APP:LzIntegral:FECG-overlap-integral}) for two FECG functions $\phi_I$ and $\phi_J$:
\begin{align}
\label{angumomfecg4.5}
\langle\hat{\bm{N}}\phi_I|\hat{\bm{N}}\phi_J\rangle &= \,\, \epsilon_{ijk}' \,\, \Bigg[ \left(\frac{\partial}{\partial \bm{e}}\omega_{I}^{(j,k)}\bm{s}_I\right) 
\left(\frac{\partial}{\partial \bm{e}}\omega_{J}^{(j,k)}\bm{s}_J\right) \Bigg] \left\langle\phi_I|\phi_J\right\rangle
\end{align}
Finally, the integral of $\hat{\bm{N}}^2$ with the $\phi_I$ and $\phi_J$ functions is obtained as
\begin{align}
\langle\hat{\bm{N}}\phi_I|\hat{\bm{N}}\phi_J\rangle = & \, \epsilon_{ijk}' \, 
\Bigg[ \left(\frac{\partial}{\partial \bm{e}}\omega_I^{(j,k)}\bm{s}_I\right) 
\left(2 \, \bm{e}^TA^{-1}\omega_J^{(j,k)}\bm{s}_J\right) \Bigg] \,\left\langle\phi_I|\phi_J\right\rangle \nonumber \\
= & \, \epsilon_{ijk}' \Bigg[ \bigg( 2\bm{s}_I^T\omega_I^{(j,k)^T}A^{-1}\omega_{J}^{(j,k)}\bm{s}_J \bigg) 
+ \bigg(2 \, \bm{e}^TA^{-1}\omega_J^{(j,k)}\bm{s}_J\bigg) \bigg( 2\, \bm{e}^TA^{-1}\omega_{I}^{(j,k)}\bm{s}_I \bigg) \Bigg]
\left\langle\phi_I|\phi_J\right\rangle
\label{eq:totAngMom_FECG_fine}
\end{align}

\subsubsection{Alternative evaluation for $\langle\phi_I|\hat{{N}}^2|\phi_J\rangle$} \label{APP:L2-FECG-alternative}

Instead of exploiting the Hermiticity of $\hat{\bm{N}}$, as in Eq.~(\ref{angumomfecg4}), 
we may directly expand $\hat{{N}}^2$ as the sum of the square of its components: 
\begin{equation}
\left\langle\phi_I\left|\hat{{N}}^2\right|\phi_J\right\rangle = \left\langle\phi_I\left|\hat{N}^2_x+\hat{N}^2_y+\hat{N}^2_z\right|\phi_J\right\rangle.
\end{equation}
Starting from Eq.~(\ref{angumomfecg2.1}), we find that by applying the operator $\hat{N}_z$ on an FECG
the action of another $\hat{N}_z$ operator produces a lengthy expression, 
\begin{align}
\hat{N}^2_z|\phi_J\rangle = -2 \epsilon_{ijk}' \sum_{P=0}^{N_p} \left(r_j^{(P)}\nabla_k^{(P)}-r_k^{(P)}\nabla_j^{(P)}\right)
\left(r_j^{(P)}A^{P(k\rightarrow j)}_{J(\text{row})}\bm{s}_J-r_k^{(P)}A^{P(j\rightarrow k)}_{J\text{(row)}}\bm{s}_J\right) \left|\phi_J\right\rangle.
\end{align}
However, after simple algebraic manipulations, the following expression is obtained:
\begin{align}
\hat{{N}}^2|\phi_J\rangle = & \epsilon_{ijk}' \left.\left. 2\left(\bm{r}A^{(j,k)}_{J(\text{row})}\bm{s}_J\right)
-2\left(\bm{r}\,\omega^{(j,k)}\bm{s}_J\right)^2 \right|\phi_J\right\rangle ~.
\end{align}
With this result, we can write the integral as 
\begin{align}
& \left\langle\phi_I\left|\hat{N}^2\right|\phi_J\right\rangle = \epsilon_{ijk}' \left[\left(\frac{\partial}{\partial\bm{e}}A^{(j,k)}_{J(\text{row})}\bm{s}_J\right)  
-\left(\frac{\partial}{\partial\bm{e}}\omega^{(j,k)}\bm{s}_J\right)^2\right] \left\langle\phi_I|\phi_J\right\rangle \nonumber \\
&= \epsilon_{ijk}' \left( 2\bm{e}^TA^{-1}A^{(j,k)}_{J{\text{(row)}}}\bm{s}_J -2\bm{s}_J\omega_J^{(j,k)^T}A^{-1}\omega_J^{(j,k)}\bm{s}_J 
+4\left(\bm{e}^TA^{-1}\omega_J^{(j,k)}\bm{s}_J\right)^2\right) \left\langle\phi_I|\phi_J\right\rangle
\end{align}
where $A^{(j,k)}_{J(\text{row})}$ is obtained from the $A_J$ matrix by setting all elements for the $i$th coordinate of every particle to zero.

The final results read
\begin{align}
\hat{\bm{N}}_{i_1}|\phi_J\rangle &= \frac{2}{i} \epsilon_{i_1j_1k_1} \left(\bm{r}^T\omega_J^{(j_1,k_1)}\bm{s}_J\right) |\phi_J\rangle,\\
\hat{\bm{N}}_{i_1}^2|\phi_J\rangle &= \epsilon_{i_1j_1k_1}' \left.\left. \left[2\left(\bm{r}A^{(j_1,k_1)}_{J(\text{row})}\bm{s}_J\right)
-2\left(\bm{r}\,\omega^{(j_1,k_1)}\bm{s}_J\right)^2\right] \right|\phi_J\right\rangle.
\end{align}

\section{Elimination of the center-of-mass contributions from the integral expressions of the square of the total angular momentum operator} 
\label{App:TI_angumom}

The center of mass (CM) of a system moves like a free particle and its states are not quantized and not square integrable. 
We eliminate the contributions from this continuous degree of freedom, $\bm{r}_{\text{CM}}$, from the angular momentum integrals
derived in Sections \ref{APP:L&LzIntegrals} and \ref{APP:L2-FECG} employing our approach described in Ref.~\cite{Muolo2018a}.

Generally speaking, the separation of space-translation variables from internal variables is a well understood problem 
\cite{Sutcliffe:coordTransf}.
The CM correction terms for the angular momentum integrals are derived in the following equations.

We start with Eq.~(\ref{eq:totAngMom_FECG_fine})
\begin{align}
\langle\hat{{N}}^2\rangle 
= \, \epsilon_{ijk}' & \Bigg[ \underset{\equiv A}{\underbrace{\bigg( 2\bm{s}_I^{(r)^T}\omega_{I}^{(j,k)^T}A^{-1}\omega_{J}^{(j,k)}\bm{s}_J^{(r)} \bigg)}}
+ \underset{\equiv B}{\underbrace{\bigg(2 \, \bm{e}^{(r)^T}A^{-1}\omega_{J}^{(j,k)}\bm{s}_J^{(r)}\bigg)}}
\underset{\equiv C}{\underbrace{\bigg( 2\, \bm{e}^{(r)^T}A^{-1}\omega_{I}^{(j,k)}\bm{s}_I^{(r)} \bigg)}} \Bigg] \nonumber \\
& \times \underset{\equiv \frac{\left\langle\phi_I|\phi_J\right\rangle}{\left(\left\langle\phi_I|\phi_I\right\rangle\left\langle\phi_J|\phi_J\right\rangle\right)^{\frac{1}{2}}}}
{\underbrace{\left(\frac{\left(\det\left(A_I\right)\det\left(A_J\right)\right)^{\frac{1}{2}}}{\det\left(A\right)}\right)^{\frac{1}{2}} 
\exp\bigg(-\bm{s}^{{(r)}T}_IA_I\bm{s}_I^{(r)}-\bm{s}_J^{{(r)}T}A_J\bm{s}_J^{(r)}+\bm{e}^TA^{-1}\bm{e}\bigg)}}
\end{align}
where $\bm{e}=A_I\bm{s}_I^{(r)}+A_J\bm{s}_J^{(r)}$, and focus on the pre-exponential terms generated by the $\hat{N}^2$ operator 
leaving aside the remaining overlap integral.
The superscript $(r)$ refer to variational vectors associated to laboratory-fixed Cartesian coordinates (LFCC). 
Superscript $(x)$ denotes 
variational vectors in transformed translationally invariant Cartesian coordinates (TICC) $\bm{s}^{(x)}$,
\begin{align}
\bm{s}^{(r)}=U_x^{-1}\bm{s}^{(x)} = U_x^{-1}\bm{s}^{(x)} \left( \begin{array}{c} {\bm{s}}' \\ \bm{c}_S \end{array} \right),
\end{align}
where $\bm{c}_S$ is a $3$-dimensional vector associated to $\bm{r}_{\text{CM}}$ and
\begin{align}
\bar{A}^{(r)}=U_{x}^T\bar{A}^{(x)}U_{x}.
\end{align}
We have
\begin{align}
A =& 2\left(\bm{s}_I^{(x)^T} {\tilde{\omega}}_I^{(j,k)^T} U_xA^{-1}U_y^{T} \tilde{\omega}_J^{(j,k)} \bm{s}_J^{(x)} \right),  \\
B =& 2\left(\bm{s}_I^{(x)^T} A_I^{(x)}U_x A^{-1} U_y^T \tilde{\omega}_J^{(j,k)} \bm{s}_J^{(x)} + 
             \bm{s}_J^{(y)^T} A_J^{(y)}U_y A^{-1} U_y^T \tilde{\omega}_J^{(j,k)} \bm{s}_J^{(x)} \right), \\
C =& 2\left(\bm{s}_I^{(x)^T} A_I^{(x)}U_x A^{-1} U_x^T \tilde{\omega}_I^{(j,k)} \bm{s}_I^{(x)} + 
             \bm{s}_J^{(y)^T} A_J^{(y)}U_y A^{-1} U_x^T \tilde{\omega}_I^{(j,k)} \bm{s}_I^{(x)} \right),
\end{align}
where $\tilde{\omega}$ indicates that it is built in the TICC set defined by the $U_x$ matrix (see also Ref.~\cite{Muolo2018a}).
Next, we recall the results of Ref.~\cite{Muolo2018a}:
\begin{align}
U_{x}\bar{A}_{IJ}^{-1}U_{y}^T &= \left[
\begin{array}{cc}
\mathcal{A}_{IJ}^{-1} & 0 \\
0 & \frac{1}{c_{A_I}+c_{A_J}}
\end{array}
\right]
\end{align}
and 
\begin{align}
U_{x}\bar{A}_{IJ}^{-1}U_{x}^T &= \left[
\begin{array}{cc}
\mathcal{A'}_{IJ}^{-1} & 0 \\
0 & \frac{1}{c_{A_I}+c_{A_J}}
\end{array}
\right],
\end{align}
to re-write $A$, $B$ and $C$ as
\begin{align}
A =\quad & 2 \left( \begin{array}{c} \bm{s}'_I \\ \bm{c}_{S_I} \end{array} \right)^T
\left[ \left( \begin{array}{cc} \mathcal{A}_I & 0 \\ 0 & c_{A_I} \end{array} \right) \otimes \mathbb{1}_{jk} \right]
\left[ \left( \begin{array}{cc} \mathcal{A}_{IJ}^{-1} & 0 \\ 0 & \frac{1}{c_{A_I}+c_{A_J}} \end{array} \right) \otimes \mathbb{1}_3 \right] \nonumber \\
& \hspace{2.25cm} \times \left[ \left( \begin{array}{cc} \mathcal{A}_J & 0 \\ 0 & c_{A_J} \end{array} \right) \otimes \mathbb{1}_{jk} \right]
\left( \begin{array}{c} \bm{s}'_J \\ \bm{c}_{S_J} \end{array} \right),
\label{APP:elimin-CM-contrib-N2-A}
\end{align}
\begin{align}
B =\quad & 2 \left( \begin{array}{c} \bm{s}'_I \\ \bm{c}_{S_I} \end{array} \right)^T
\left[ \left( \begin{array}{cc} \mathcal{A}_I & 0 \\ 0 & c_{A_I} \end{array} \right) \otimes \mathbb{1}_3 \right]
\left[ \left( \begin{array}{cc} \mathcal{A}_{IJ}^{-1} & 0 \\ 0 & \frac{1}{c_{A_I}+c_{A_J}} \end{array} \right) \otimes \mathbb{1}_3 \right] \nonumber \\
& \hspace{2.25cm} \times \left[ \left( \begin{array}{cc} \mathcal{A}_J & 0 \\ 0 & c_{A_J} \end{array} \right) \otimes \mathbb{1}_{jk} \right]
\left( \begin{array}{c} \bm{s}'_J \\ \bm{c}_{S_J} \end{array} \right) \nonumber \\
+& 2 \left( \begin{array}{c} \bm{s}'_J \\ \bm{c}_{S_J} \end{array} \right)^T
\left[ \left( \begin{array}{cc} \mathcal{A}_J & 0 \\ 0 & c_{A_J} \end{array} \right) \otimes \mathbb{1}_3 \right]
\left[ \left( \begin{array}{cc} \mathcal{A''}_{IJ}^{-1} & 0 \\ 0 & \frac{1}{c_{A_I}+c_{A_J}} \end{array} \right) \otimes \mathbb{1}_3 \right] \nonumber \\
& \hspace{2.25cm} \times \left[ \left( \begin{array}{cc} \mathcal{A}_J & 0 \\ 0 & c_{A_J} \end{array} \right) \otimes \mathbb{1}_{jk} \right]
\left( \begin{array}{c} \bm{s}'_J \\ \bm{c}_{S_J} \end{array} \right),
\label{APP:elimin-CM-contrib-N2-B}
\end{align}
\begin{align}
C =\quad & 2 \left( \begin{array}{c} \bm{s}'_I \\ \bm{c}_{S_I} \end{array} \right)^T
\left[ \left( \begin{array}{cc} \mathcal{A}_I & 0 \\ 0 & c_{A_I} \end{array} \right) \otimes \mathbb{1}_3 \right]
\left[ \left( \begin{array}{cc} \mathcal{A'}_{IJ}^{-1} & 0 \\ 0 & \frac{1}{c_{A_I}+c_{A_J}} \end{array} \right) \otimes \mathbb{1}_3 \right] \nonumber \\
& \hspace{2.25cm} \times \left[ \left( \begin{array}{cc} \mathcal{A}_I & 0 \\ 0 & c_{A_I} \end{array} \right) \otimes \mathbb{1}_{jk} \right]
\left( \begin{array}{c} \bm{s}'_I \\ \bm{c}_{S_I} \end{array} \right) \nonumber \\
+& 2 \left( \begin{array}{c} \bm{s}'_J \\ \bm{c}_{S_J} \end{array} \right)^T
\left[ \left( \begin{array}{cc} \mathcal{A}_J & 0 \\ 0 & c_{A_J} \end{array} \right) \otimes \mathbb{1}_3 \right]
\left[ \left( \begin{array}{cc} \mathcal{A}_{IJ}^{-1} & 0 \\ 0 & \frac{1}{c_{A_I}+c_{A_J}} \end{array} \right) \otimes \mathbb{1}_3 \right] \nonumber \\
& \hspace{2.25cm} \times \left[ \left( \begin{array}{cc} \mathcal{A}_I & 0 \\ 0 & c_{A_I} \end{array} \right) \otimes \mathbb{1}_{jk} \right]
\left( \begin{array}{c} \bm{s}'_I \\ \bm{c}_{S_I} \end{array} \right),
\label{APP:elimin-CM-contrib-N2-C}
\end{align}
where $\mathbb{1}_{jk}=E_{kj}-E_{jk}$ and $E_{ij}$ is a $3\times3$ matrix in which only the $ij$-th element is different from zero
and equal to $1$.

Following the prescriptions of Ref.~\cite{Muolo2018a}, the CM contributions are eliminated by subtracting the following terms 
from A, B, and C in Eqs.~(\ref{APP:elimin-CM-contrib-N2-A})--(\ref{APP:elimin-CM-contrib-N2-C}):
\begin{align}
A^{\text{(CM)}} =& \,2\, \bm{c}_{S_I}^T \left(c_{A_I}\otimes\mathbb{1}_{jk}\right) \left(\frac{1}{c_{A_I}+c_{A_J}}\otimes\mathbb{1}_3\right)
                     \left(c_{A_J}\otimes\mathbb{1}_{jk}\right) \bm{c}_{S_J},
\end{align}
\begin{align}
B^{\text{(CM)}} =& \,2\, \bm{c}_{S_I}^T \left(c_{A_I}\otimes\mathbb{1}_{3}\right) \left(\frac{1}{c_{A_I}+c_{A_J}}\otimes\mathbb{1}_3\right)
                     \left(c_{A_J}\otimes\mathbb{1}_{jk}\right) \bm{c}_{S_J}  \nonumber \\
                 & \,+2\, \bm{c}_{S_J}^T \left(c_{A_J}\otimes\mathbb{1}_{3}\right) \left(\frac{1}{c_{A_I}+c_{A_J}}\otimes\mathbb{1}_3\right)
                     \left(c_{A_J}\otimes\mathbb{1}_{jk}\right) \bm{c}_{S_J},
\end{align}
\begin{align}
C^{\text{(CM)}} =& \,2\, \bm{c}_{S_I}^T \left(c_{A_I}\otimes\mathbb{1}_{3}\right) \left(\frac{1}{c_{A_I}+c_{A_J}}\otimes\mathbb{1}_3\right)
                     \left(c_{A_I}\otimes\mathbb{1}_{jk}\right) \bm{c}_{S_I}  \nonumber \\
                 & \,+2\, \bm{c}_{S_J}^T \left(c_{A_J}\otimes\mathbb{1}_{3}\right) \left(\frac{1}{c_{A_I}+c_{A_J}}\otimes\mathbb{1}_3\right)
                     \left(c_{A_I}\otimes\mathbb{1}_{jk}\right) \bm{c}_{S_I}.
\end{align}

\section{Alternative projection approach} \label{SEC:alternative_proj_approach}

L\"owdin proposed two different forms for the angular momentum projection: a sum \cite{newproj_L_1,newproj_L_2}
which is derived from a product form \cite{newproj_L_0}. 
His original method views the projector as a product of annihilation operators that remove 
all components other than that of the desired symmetry.
This iterative process is accomplished by means of the two operators $\hat{P}_N$ and $\hat{P}_{M_N}$ 
acting on an arbitrary function $\psi$ that is resolved into components $C_{NM_N}\psi_{NM_N}$
which are eigenfunctions of $\hat{N}^2$ and $\hat{N}_z$:
\begin{align}
\label{Lowproj1}
\psi = \sum_{N}\sum_{M_N} C_{NM_N}\psi_{NM_N} ~,
\end{align}
where the summation is over all possible values of $N$ and $M_N$ and the wavefunction is written in the basis
of mutually orthogonal unit vectors $\psi_{NM_N}$ spanning the complete Hilbert space.
The eigenvalue relations for $\hat{N}^2$ and $\hat{N}_z$ may be written in the form
\begin{align}
\label{Lowproj2}
\left[\hat{N}^2-N(N+1)\right]\psi_{NM_N} &\equiv 0 \\
\label{lowproj3}
\left[\hat{N}_z-M_N\right]\psi_{NM_N} &\equiv 0
\end{align}
which means that the eigenfunction $\psi_{NM_N}$ is annihilated by the operator $\left[\hat{N}^2-N(N+1)\right]$
or $\left[\hat{N}_z-M_N\right]$. 
It is therefore possible to select a specific component, $C_{NM_N}\psi_{NM_N}$, from $\psi$
by annihilating all other components. This can be achieved with the following projectors \cite{newproj_L_0}
\begin{align}
\label{Lowproj3}
\hat{P}_N &= \prod_{l\ne N} \frac{\hat{N}^2-l(l+1)}{N(N+1)-l(l+1)}, \\
\label{Lowproj4}
\hat{P}_{M_N} &= \prod_{\mu\ne M_N} \frac{\hat{M}_N-\mu}{M_N-\mu},
\end{align}
where the numerators are products of L\"owdin's elementary annihilation operators over all quantum numbers except 
those which correspond to the selected value(s). 
The denominators have been chosen so that the projectors have eigenvalue $1$ when acting on the term $\psi_{NM_N}$.

L\"owdin showed that Eq.~(\ref{Lowproj3}) may be rewritten, when acting on an eigenvector of $N_z$ 
with eigenvalue $M_N\ge0$, in the so-called sum form \cite{newproj_L_1}
\begin{align}
\label{Lowproj5}
\hat{P}_{NM_N} &= \prod_{l\ne N} \frac{(2N+1)(N+M_N)!}{(N-M_N)!} \sum_{l\ne N} 
\frac{(-1)^l N_{-}^{N-M_N+k}N_{+}^{N-M_N+k}}{k!(2N+1+k)!}
\end{align}
where 
\begin{align}
N_{\pm} = N_x \pm iN_y
\end{align}
are the usual raising and lowering operators and the subscript $M_N$ is added to $P_N$ since Eq.~(\ref{Lowproj5})
is valid only when acting on a state of definite $M_N$ (FECG functions have $M_N=0$).

For practical applications the product operators (\ref{Lowproj3}) and (\ref{Lowproj4}) can be restricted to contain 
only a finite number of factors, $l_{\text{max}}$. 
However, the product series in Eq.~(\ref{Lowproj3}) converges quadratically with respect to $l$ since the $l$-th term is $\approx 1$ for sufficiently large $l$
\cite{newproj_L_0}.

We have not explored the feasibility of a numerical approach based on Eq.~(\ref{Lowproj3}) or Eq.~(\ref{Lowproj5})
because the expressions for $\hat{N}_i^n$, for $n$ being a positive integer, become lengthy 
as shown in Appendix \ref{APP:L2-FECG-alternative} already for $n=2$.

\end{appendix}

\newcommand{\Aa}[0]{Aa}

\end{document}